\documentclass[usegraphicx,usenatbib]{mn2e}

\usepackage{natbib,graphicx}

\usepackage{epsfig}
\usepackage{ulem}
\usepackage{color}
\usepackage{graphics,graphicx}
\usepackage{times} 
\usepackage{amssymb}
\usepackage{amsmath}
\usepackage{url}
\usepackage{multirow}
\usepackage{ctable}
\usepackage{threeparttable}
\newif\ifAMStwofonts
\AMStwofontstrue
\def\mekal{\rm{\sc MEKAL}}
\def\wabs{\rm{\sc WABS}}
\def\nh{${\it N}_{\rm H}$}

\def\asca{{\it ASCA}}
\def\chandra{{\it Chandra}}

\def\rosat{{\it ROSAT}}

\def\xspec{\hbox{\sc XSPEC}}
\def\xspecv{{\sc XSPEC}{\rm\thinspace v\thinspace 12.7.0}}

\def\xselect{{\sc XSELECT}}
\def\grppha{{\sc GRPPHA}}

\def\cflux{\rm{\sc CFLUX}}

\def\cm{\hbox{$\rm\thinspace cm$}}
\def\pc{\hbox{$\rm\thinspace pc$}}
\def\kpc{\hbox{$\rm\thinspace kpc$}}

\def\degsq{\hbox{$\rm\thinspace deg^{2}$}}

\def\pcmsq{\hbox{$\rm\thinspace cm^{-2}$}}

\def\kev{\hbox{$\rm\thinspace keV$}}

\def\ergpcmsqps{\hbox{$\rm\thinspace erg~cm^{-2}~s^{-1}$}}

\def\ergps{\hbox{$\rm\thinspace erg~s^{-1}$}}

\def\msun{\hbox{$\rm\thinspace M_{\odot}$}}

\def\k{\hbox{\rm\thinspace K}}
\def\kt{\rm kT}

\def\swift{{\it Swift}}

\begin{document}

\title[The Luminosity Function of CVs and the GRXE] {X-ray Luminosities of Optically-Selected  Cataclysmic Variables and Application to the Galactic Ridge X-ray Emission}
\author[Reis et al.]{
   R.~C.~Reis $^{1}$\thanks{E-mail: rdosreis@umich.edu},
   P.~J.~Wheatley${^2}$\thanks{E-mail: p.j.wheatley@warwick.ac.uk},
   B.~T.~G\"ansicke${^2}$, \&
   J.~P.~Osborne$^3$
\medskip\\
$^{1}$Institute of Astronomy, Madingley Road, Cambridge, CB3 0HA\\
$^{2}$Department of Physics, University of Warwick, Coventry, CV4 7AL\\
$^3$University of Leicester, University Road, Leicester, LE1 7RH\\
}
\maketitle

\begin{abstract}  
By studying \swift\ X-ray spectra of an optically-selected, non-magnetic sample of nearby cataclysmic variables (CVs), we show that there is a population with X-ray luminosity much lower than accounted for in existing studies. We find an average 0.5-10.0\kev\ luminosity of $8\times10^{29}\ergps$ which is an order of magnitude lower than observed in previous samples. Looking at the co-added X-ray spectrum of twenty CVs, we show that the spectral properties of this optically-selected, low X-ray luminosity sample -- likely characteristic of the dominant population of CVs -- resembles that of their brighter counterpart, as well as the  X-ray emission originating in the Galactic ridge. It is argued that if the space density of CVs is greater than the current  estimates, as it is indeed predicted by population synthesis models, then CVs can significantly contribute to the Galactic ridge emission.
\end{abstract}


\section{Introduction}
\label{s-intro}
The serendipitous detection in 1962 of an almost uniform Cosmic X-ray Background (CXB) is regarded as one of the first discoveries of extrasolar X-ray astronomy \citep{Giacconi1962} and is thought to be mostly due to unresolved active galactic nuclei together with type~Ia supernovae \citep[e.g.][]{zdziarski96, DraperBallantyne2009, Moretti2009}.  A decade later the possibility of a Galactic component to the total measured CXB was suggested \citep*{Cooke70}. Ensuing observations showed that this Galactic component to the X-ray background is concentrated near the Galactic disk, covering a continuous ridge of
width $\sim10 \deg.$ extending along the Galactic plane for tens of degrees -- the Galactic Ridge X-ray Emission (GRXE, see e.g. \citealt{Worrall1982, Warwick1985, Warwick1988}).

Numerous authors have modelled the GRXE spectrum as originating from two collisionally-ionised plasma components (e.g., \citealt{Koyama1986, Koyama1996, Kaneda1997}): A soft component (\kt$ \sim0.8$\kev) possibly produced by supernova shock waves \citep[e.g.][]{Kaneda1997}  and a hard (\kt$ \sim8$ \kev) component whose origin is still a topic of much discussion. Several explanations for the nature of this hard 
component have been suggested, amongst them: {\rm i)}  a superposition of discrete, faint X-ray sources such as cataclysmic variables
\citep{Worrall1982, Revnivtsev2006, Revnivtsev2009} or {\rm ii)} originating from diffuse gas regions such as molecular clouds
illuminated by X-rays from point sources (e.g. \citealt{Murakami2000}) or bombarded by low-energy cosmic-ray electrons (e.g. \citealt{Yusef2002}), and/or non-thermal processes in the interstellar medium (e.g. \citealt{Dogiel2002}). The apparent thermal spectrum of the GRXE
\citep{Koyama1986} implies an emitting plasma with temperature of a few \kev. The energy density of this hypothetical plasma
is over two orders of magnitude higher than that of normal interstellar matter and its temperature is too high for it to be
gravitationally bound to the Galactic plane \citep{Koyama1996}. Furthermore \citet{Dogiel2002} showed that the energy
required to replenish the outflowing plasma is as high as $10^{42} \ergps$ which is equivalent to the release of kinetic energy from one
supernova occurring every 30 years in the Galactic plane region. With an expected rate of 2-3 supernovae per century in
the whole Galaxy \citep{Tammann1982}, this leads to the unlikely requirement that all the energy associated with  supernovae in the Galaxy   goes into exclusively  heating this hypothetical diffuse plasma.

The absence of a plausible heating mechanism to this hypothetical
diffuse thermal plasma, and the similarities of the X-ray spectra of
the ridge and CVs, led many to believe that the
ridge emission is still most likely to arise from as yet unresolved sources such as 
cataclysmic variables (CVs) and other accreting binary systems \citep{Muno2004, Revnivtsev2007PThPS}.  An ultra-deep (1~Ms) Chandra ACIS-I  study of the Galactic centre region \citep{Revnivtsev2009} has successfully resolved over 80~percent of the diffuse emission into discrete sources at 6--7\kev\ (see also \citealt{Hong2012,Yuasa2012}), however the  nature of these sources remains uncertain \citep[e.g.][]{Warwick2011} and the contribution of CVs to the total GRXE luminosity function is still known. A key uncertainty is the X-ray luminosity function of CVs. Previous studies were suggestive of mean CV X-ray luminosities of log$(<L_{\rm x}>)\sim 31.5$ \citep{Mukai1993, Verbunt1997} in the hard band (2.0--10.0\kev), with at least an order of magnitude spread about this mean value. Work done on \asca\ data \citep{Yamauchi1996} concluded that for CVs to be accountable for the GRXE the sources' 2.0--10.0\kev\ luminosity could not be higher than $2\times10^{33}\ergps$. Based on model calculations, \citet{Tanaka1999} showed that the ridge emission requires a class of hard sources with a luminosity in the $10^{29-30}\ergps$ range, thus making CVs apparently too luminous as a class.

\citet{Sazonov2006} found CVs contributing only at $L_{\rm x}\gtrsim10^{31}\ergps$,  but their sample was X-ray selected and were mostly consistent of magnetic CVs, and thus biased to high X-ray luminosities. On the other hand, a more recent study by \citet{Byckling2010} used a  distance-limited sample of  non-magnetic CVs with known distances and found systematically low luminosities concentrated around $L_{\rm x}\approx 10^{30.5}\ergps$. However, even this sample was biased to high accretion rates by reliance on inhomogeneous selection criteria such as dwarf nova outburst.

Until recently, the population of known CVs has been dominated
  by systems discovered because they displayed copious accretion
  activity, either in the form of outbursts, or X-ray emission
  \citep{gaensickeetal02, gaensicke05}. In contrast to this,
  population synthesis models have long predicted that the CV
  population should be dominated ($>$95\%) by short-period
  low-luminosity systems with extremely low-mass donor stars
  \citep{Kolb1993, Howell2001} and thus should have  average X-ray luminosities
   $\lesssim10^{30} \ergps$. Over the past years, the Sloan Digital Sky
  Survey (SDSS) has identified a substantial number of CVs whose
  optical spectra are dominated by a cool white dwarf
  \citep[e.g.][]{Szkody2009}, implying very low secular average
  accretion rates \citep{Townsley2003, Townsley2009, gaensickeetal09}, and that have extremely low-mass donor stars \citep{Littlefair2006Sci, littlefairetal08-1}. This sample of optically-selected, non-magnetic CVs provide the means to measure the 
  X-ray luminosity of CVs towards the low-end of their luminosity function and thus to  better assess their contribution to the Galactic ridge X-ray
  emission.

In the following sections we present results on \swift-XRT
observations made on a sample of 20 CVs whose optical spectra are
dominated by the white dwarf, indicative of low accretion
rates. These  are the systems expected to dominate in numbers the population of CVs.

\section{Target sample}
\label{s-sample}

The Sloan Digital Sky Survey (SDSS; \citealt{sdss}) is
  an imaging and spectroscopic survey of the high Galactic latitude
  sky visible from the Northern hemisphere. We based our target
  selection on Data Release\,5 (DR5), which covered over 8,000
  deg$^2$ of the sky and performed spectroscopy on over 1\,million
  objects \citep{Adelman-McCarth2006}. Although the principle
  goal of this survey is to survey the large-scale distribution of
  galaxies and quasars, it has also provided vast amount of data on
  stars.  Due to CVs having non-stellar colours they are
  serendipitously discovered by SDSS. \citet{Szkody2011} have identified close to 300 CVs, of which
  $\sim45$ have a very low mass accretion rate and are thus dominated
  by the white dwarf.

For the \textit{Swift} survey presented here, we selected 16
  CVs from SDSS DR5 in which the optical spectrum is dominated by
  emission from a cool white dwarf. These spectra have been sky
  subtracted, corrected for telluric absorption and
  spectrophotometrically calibrated by the Spectro2d pipeline
  \citep{Stoughton2002}. This sample was supplemented by four additional
  low-mass transfer systems with similar optical properties:
  V455\,And \citep{Araujo-Betancor2005}, PQ\,And
  \citep{Schwarz2004, Patterson2005}, RE\,J1255+266
  \citep{Watson1996, Wheatley2000, Patterson2005P} and ASAS0025+1217 (aka  FL Psc,  \citealt{templetonetal06-1, ishiokaetal07-1}). The
  object names and coordinates of the 20 \textit{Swift} targets are
  listed in Table~1. Our targets form a representative sample of
  spectroscopically selected low-luminosity CVs with distances in the
  range $\sim75-400$\,\pc\ (see Sect.\,~\ref{s-distances}).

\section{Observations and Data reduction}
\label{s-observations}

Our sample of 20 optically-selected, non-magnetic CVs were observed using the
X-Ray Telescope, XRT (0.2-10.0 keV) on board the \swift\ Gamma-Ray
Burst Explorer \citep{swift}. The various observations, listed
in Table~1, were reduced using the tools provided in the HEASOFT
v{\thinspace 6.9} software package. All data were extracted in the
photon counting (PC) mode with the standard grade
selection (0--12) for this mode. For each observation an image in
the 0.2-10.0\kev\ band was obtained from which a spectrum was
extracted from a circular region of radius 47'', corresponding to 90
percent of the point spread function at 1.5\kev. Background spectra were extracted from an annular region centred on the X-ray source of inner and outer radius of 50'' and 200'' respectively. Response files (version 10) were downloaded{\footnote{http://heasarc.nasa.gov/docs/heasarc/caldb/data/swift/xrt/index.html}} and used for
spectral analysis. When looking at the spectrum of each individual
target we used the {\rm ftool}{\footnote{http://heasarc.gsfc.nasa.gov/ftools/} \grppha\ to provide a minimum of 4 counts per bin and thus provide two or
  more independent spectral bins. Following this requirement only
  sources with 8 or more counts were individually analysed in this
  manner. Model fits were minimised using Cash statistics (C-stat in
  \xspec; \citealt{Cash1979}) due to the low number of counts per bin. All
  spectral analyses were performed using \xspecv\ \citep{XSPEC}.

\begin{table*}
\setlength{\tabcolsep}{0.7ex}
\begin{center}
\caption{System properties \label{t-results}}
\begin{tabular}{lcccccccccccccc}                
\hline
\hline

Target & RA(2000) & Dec(2000)   &   $P_\mathrm{orb}$& $g$  &Live&  Counts & Model Flux$^{a}$ & Distance$^{b}$ & Distance$^{c}$  & Luminosity  & References \\
       &    &    &    &                &    time (s)  & [0.5--10\kev]  &$\times10^{-13}$            &                &           & [0.5--10.0\kev]  \\
       &        && (min)           &    mag.  &  && $\ergpcmsqps$ &  (pc)          &  (pc)     &  $\times10^{29}$(erg s$^{-1}$) \\
\hline
SDSS0131--0901&  01 31 32.39 &-09 01 22.3  & 81.5  & 18.3& 4940  	& 23.9       & $2.0\pm0.5$  &   240        & 219$\pm$72 &   $13.8\pm6.4 $     & 1   \\
SDSS0137--0912& 01 37 01.06&-09 12 34.9   & 79.7  & 18.7 & 4493  	& 16.2       & $1.5\pm0.4 $ &   230        & 263$\pm$86 &   $9.4\pm4.6 $  & 2   \\
SDSS0843+2751 &08 43 03.99&+27 51 49.7   & 85.5  & 18.9 &476     	&$<1^d$       & $<0.3$       &   240        & 288$\pm$95 &   $ <1.8$       & 3,4 \\
SDSS0904+0355 &  09 04 03.48&+03 55 01.2   &  86.0  & 19.3 &5670	& 16.8     & $1.2\pm0.3 $ &   260        & 347$\pm$114&   $ 9.9\pm4.8$  & 5   \\
SDSS0904+4402 &  09 04 52.09&+44 02 55.4   &     & 19.4 &4367		& $1.0$      & $0.09\pm0.01$&  320        & 363$\pm$119&    $ 1.2\pm0.5$  &  6   \\
SDSS0919+0857 &09 19 45.11&+08 57 10.0   &   81.3  & 18.2 &167		&$<1^d$       & $ <2$        &  220        & 209$\pm$69 &    $ <11$        & 6,7 \\
SDSS1137+0148 &11 37 22.25&+01 48 58.6   & 109.6 & 18.7 &3798	& 11.8      & $1.3\pm0.4 $ &   220        & 263$\pm$86 &   $7.4\pm0.4 $  & 8   \\
SDSS1238--0339& 12 38 13.73&-03 39 33.0   &80.5  & 17.8 &4106		& 40.3       & $4.1\pm0.8 $ &   180        & 174$\pm$57 &   $15.7\pm7.0 $  & 9   \\
SDSS1339+4847 &  13 39 47.12&+48 47 27.5   &82.5  & 17.7 &4051	& 16.7       & $1.7\pm0.5$  &   170        & 166$\pm$54 &   $ 5.9\pm2.8$  & 10  \\
SDSS1457+5148 &   14 57 58.21&+51 48 07.9   &  77.92  & 19.6 &6293		& 14.8       & $ 1.0\pm0.3$ &   320        & 398$\pm$131&   $ 11.9\pm5.8$   & 11 \\
SDSS1501+5501 & 15 01 37.22&+55 01 23.4   &81.9  & 19.4&4800 		& 11.6      & $1.0\pm0.3 $ &   300        & 363$\pm$119&   $ 10.7\pm5.5$   & 12\\
SDSS1507+5230 & 15 07 22.33&+52 30 39.8   & 66.6  & 18.3 & 7002 	& 15.6       & $0.9\pm0.3 $    &   225        & 219$\pm$72 &  $ 5.6\pm2.7$   & 12,13,14\\
SDSS1556--0009& 15 56 44.24&-00 09 50.2   & 106.7 & 18.0 & 7463 	& 42.8       & $2.4\pm0.4 $  &   135        & 191$\pm$63 &   $5.2\pm2.3 $  & 7,15  \\
SDSS1610--0102& 16 10 33.64&-01 02 23.3   &  80.5  & 19.0&9400  	& 18.8       & $0.8\pm0.2 $ &   240        & 302$\pm$99 &  $5.7\pm2.7 $  & 16,17\\
SDSS1702+3229 &17 02 13.26&+32 29 54.1   & 144.1 & 17.9 &6840  	& 25.7       & $1.5\pm0.4 $ &   180        & -          &   $6.0\pm2.8 $ & 18,19 \\
SDSS2048--0610&  20 48 17.85&-06 10 44.8   & 87.3  & 19.4& 14734 & 68.6       & $1.9\pm0.3 $ &   270        & 363$\pm$119&  $ 16.8\pm7.3$   & 20,21 \\
ASAS0025+1217 & 00 25 11.07&+12 17 12.1   & $\simeq80$ & 17.4$^v$ &3309 	& 53.8& $6.7\pm1.2$ & 125    & 145$\pm$47         &  $ 12.6\pm5.5 $ & 22,23   \\
PQ\,AND       & 02 29 29.54&+40 02 40.2   &  $\simeq80$ & 19.0$^v$ &13215 		& 16.5          & $0.5\pm0.1 $ & $150\pm50$ & 302$\pm$99 &  $1.4\pm1.0$ &24,25 \\
RE1255+266    & 12 55 10.56&+26 42 26.9   & $\simeq120$?& 19.2&4298  			&  8.8           & $ 0.8\pm0.3$ & $180\pm50$ & 326$\pm$107&  $3.3\pm2.2$ &26,27 \\
V455\,And     & 23 34 01.45&+39 21 41.0   & 81.1  & 16.5$^v$&8666 						& 16.0		& $0.8\pm0.2$   &  $90\pm15$    & 95$\pm$31  &  $ 0.7\pm0.3  $& 28   \\
\hline
\hline
\end{tabular}
\end{center}
\parbox{\textwidth}{
\small 
\textbf{Notes:}
$^{a}$ Fluxes were calculated assuming the flux--count rate
relationship given in the text (see Sect.~\ref{s-xrayfluxes}) in the 0.5--10~keV range.
$^{b}$ distances obtained by modelling the optical SDSS spectra, as
described in the Sect.\,\ref{s-distances}, with uncertainties of 20\%,
except for the last four targets, where published distance estimates
are given. The distances listed in this column are used in the X-ray analysis. 
$^{c}$ distances obtained from $<M_g>=11.6\pm0.7$
(Gaensicke et al. 2009), except for SDSS\,1702+3229, whose period is
too long to adopt this absolute magnitude.
$^{v}$ $V$-band magnitude.
$^{d}$ $1\sigma$ upper limit.
\textbf{References:}
$^{1}$~\citet{southworthetal07-2};      % 0131-090
$^{2}$~\citet{pretoriusetal04-1};       % 0137-0912
$^{3}$~\citet{pattersonetal98-4};       % EG Cnc
$^{4}$~\citet{katoetal04-1};            % EG Cnc
$^{5}$~\citet{woudtetal05-1};           % 0904+0355
$^{6}$~\citet{dillonetal08-1};          % 0919+0857
$^{7}$~Thorstensen et al. in prep;      % 0919+0857
$^{8}$~\citet{pattersonetal03-1};       % RZ Leo
$^{9}$~\citet{zharikovetal06-1};        % 1238-0339
$^{10}$~\citet{gaensickeetal06};      % 1339+4847
$^{11}$~\citet{Uthas2012}; %1457+5148
$^{12}$~\citet{littlefairetal08-1};     % 1501+5501, 1507+5230,
$^{13}$~\citet{littlefairetal07-1};     % 1507+5230
$^{14}$~\citet{pattersonetal08-1};      % 1507+5230
$^{15}$~\citet{woudtetal04-1};          % 1556-0009
$^{16}$~\citet{woudt+warner04-1};       % 1610-0102
$^{17}$~\citet{copperwheatetal09-1};    % 1610-0102
$^{18}$~\citet{littlefairetal06-1};      % 1702+3229
$^{19}$~\citet{boydetal06-2};            % 1702+3229
$^{20}$~\citet{woudtetal05-1};           % 2048-0610
$^{21}$~Dillon et al. in prep;           % 2048-0610
$^{22}$~\citet{templetonetal06-1};       % ASAS0025
$^{23}$~\citet{ishiokaetal07-1};         % ASAS0025
$^{24}$~\citet{pattersonetal05-1};       % PQ And
$^{25}$~\citet{vanlandinghametal05-1};   % PQ And
$^{26}$~\citet{wheatleyetal00-1};        % PQ And
$^{27}$~\citet{pattersonetal05-2};       % PQ And
$^{28}$~\citet{Araujo-Betancor2005} % V455 And

}
\end{table*}

The source and background event files were used to calculate
count-rates, {\it R}, for each target in the 0.5-10.0\kev\ energy
band. The net counts were computed from the total counts in the
source region less the background count, scaled by the ratio of the
source/background area. The count-rate for each individual target
was computed using the total live-time and the uncertainties are
assumed to follow simple Poisson statistics. Table\,\ref{t-results} list the net counts for the various observations. In order to analyse the
sample as a group, in Sect.\ref{s-xrayspectra} we co-added all the individual
spectra using the {\rm ftool} \xselect. For the co-added spectrum,
a minimum of 20 counts per bin were used and model fits were
minimised using $\chi^{2}$ statistics.

\begin{figure*}
\includegraphics[angle=-90, width=8.5cm] {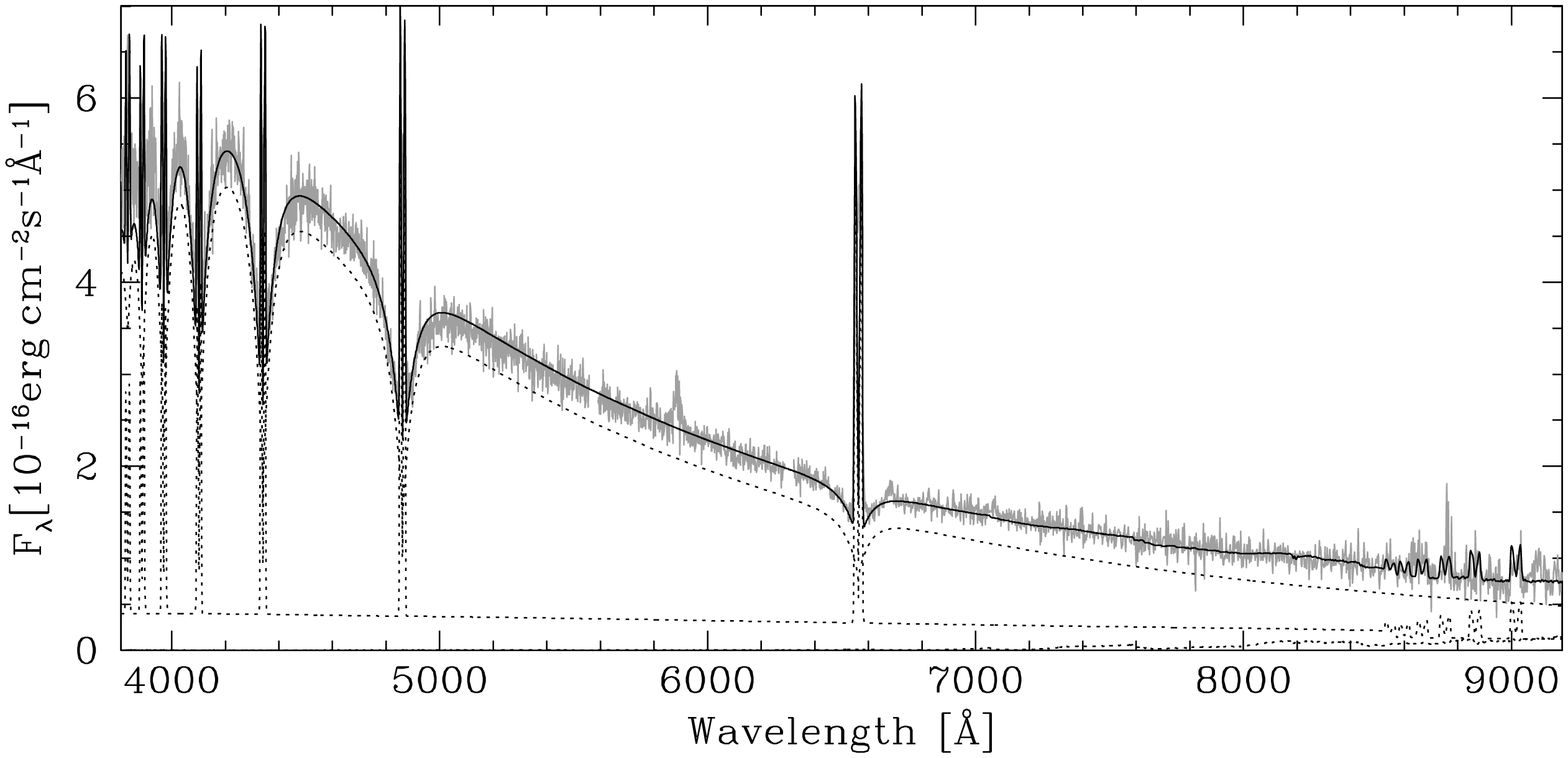}
\includegraphics[angle=-90, width=8.5cm] {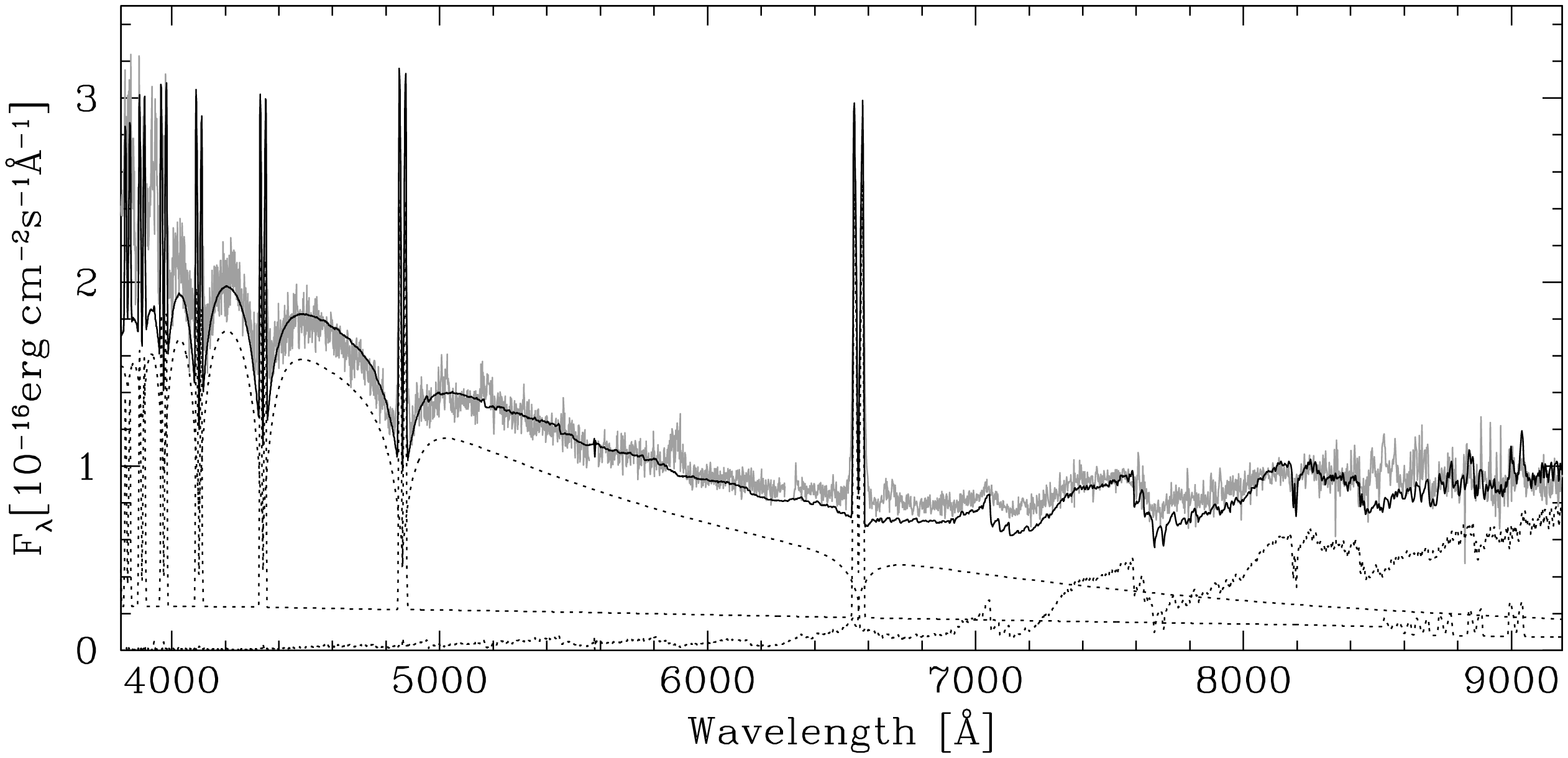}
\caption{\label{f-optfit} Three components model (white dwarf,
  optically thin disc, secondary star) of the average optical spectrum
  of SDSS\,1339+4842 (\textit{left}, Sp(2)\,=\,M8, $T_\mathrm{wd}=12500\k$,
  $T_\mathrm{d}=6500\k$, $\Sigma_\mathrm{d}=1.7\times10^{-2}{\rm
    g}{\rm cm}^{-2}$ and $d=170\pc$) and SDSS\,1137+0148 (\textit{right},
  Sp(2)\,=\,M6.5, $T_\mathrm{wd}=13000\k$, $T_\mathrm{d}= 6500\k$,
  $\Sigma_\mathrm{d}=3.5\times10^{-2}{\rm g}{\rm cm}^{-2}$ and
  $d=220\pc$). The observed data are shown in grey, the individual
  three components as dotted black lines, and the summed model as
  solid black line. }
\end{figure*}

\section{Analysis and Results}

\subsection{Distance estimates}
\label{s-distances}

In order to calculate the luminosity distribution of the sample it was
necessary to obtain an approximate distance to each individual
source. This was done following two different approaches.

\begin{figure*}
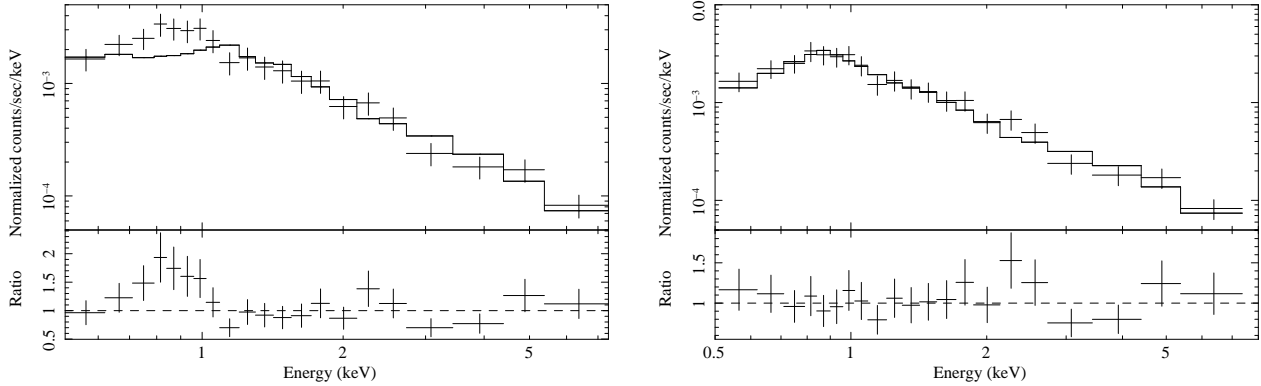

{\includegraphics[height=8.cm, angle=270] {figure6_1mekal.eps}}
\hspace{0.5cm}
{\includegraphics[height=8.cm, angle=270 ] {figure6_part2.eps}}
\caption{\label{f-spec}\swift-XRT X-ray background subtracted spectrum of combined
  sources. {\it Left}: Data/model of simple one-temperature thermal plasma
  with a temperature of $\sim6\kev$. This model provided a
  poor fit with $\chi^2/\nu =31.3/19$ and exhibited residual at
  energies $\la 1.2\kev$. {\it Right}: Data/model ratio for
  two-temperature thermal plasma with temperatures of $8^{+10}_{-3}$
  and $0.62^{+0.16}_{-0.24}$\kev\ as well as photoelectric
  absorption. This provided the best fit ($\chi^{2}/\nu=11.3/17$) for
  the combined data.}
\end{figure*} 

The first method consists of fitting a three-component model
accounting for the flux contributions of the white dwarf, the
accretion disc, and the companion star to the optical spectra of the
16 CVs for which SDSS spectroscopy is available. A detailed
description of this approach can be found in \citet{gaensickeetal97,
  gaensickeetal99, gaensickeetal06}, here we provide only a brief
summary. The white dwarf is represented by model spectra computed with
the TLUSTY/SYNSPEC codes of \citet{lanz+hubeny95}, with a fixed
radius $R_{\rm wd} = 8.7\times 10^{8}\cm$ implied by the
\citet{hamada+salpeter61} mass-radius relation for an assumed mass
of $M_{\rm wd} = 0.6\msun$. The secondary star is represented by
observed templates covering spectral types (Sp(2)) M0.5 to M9 from
\citet{beuermannetal98} and L0 to L8 from
\citet{kirkpatricketal00, kirkpatricketal99}. The radius of the
secondary star is estimated from Roche-lobe geometry, and depends
primarily on the orbital period. Orbital period measurements are
available for most of the targets (Table~1), and it is
apparent that most of the systems are located close to the 80\,min
period minimum, for which we adopt $R_{2} = 8.6\times10^9 \cm$. For
SDSS\,1137+0148 (RZ\,Leo), SDSS\,1556--0009, and SDSS\,1702+3229, we
assume $1.28\times10^{10}\,\cm$, $1.17\times10^{10}\,\cm$, and
$1.53\times10^{10}\,\cm$, respectively. Free parameters in the
three-component model are the white dwarf temperature $T_\mathrm{wd}$,
the distance $d$, the temperature $T_\mathrm{d}$ and column density
$\Sigma_\mathrm{d}$ of the disk, as well as the spectral type of the
secondary star. The parameters were varied in a forward-modelling
approach until all three spectral components in the observed spectrum
were consistently reproduced, Fig.~\ref{f-optfit} shows two example
fits (SDSS\,1137+0148 and SDSS1337+0148), and Table\,\ref{t-results}
lists the distances determined by these fits. The uncertainty of the
distances determined in this way is estimated to be $\simeq20$\% \citep[see][for a detailed description of the method employed here and associated uncertainties.]{Howell2002ApJ,Szkody2002Apj, gansicke2005ApJ, gaensickeetal06b, gaensickeetal06}.

Four of our targets have no SDSS spectrum, and Table~1
lists previously published distance estimates and references.

As a second method of estimating the distances to our targets we
adopted the mean absolute magnitude for short-period CVs with white
dwarf dominated spectra, $<M_g>=11.6\pm0.7$, which \citet{gaensickeetal09} established largely based on systems for
which ultraviolet spectroscopy is available, and which leads to a $\sim30\%$ uncertainty in the distance. The distances
corresponding to the observed $g$ and $V$ magnitudes are also listed
in Table~1, with the exception of SDSS\,1702+3229,
whose orbital period is too long to qualify for this method. For all
systems,  both distance estimates agree within the quoted errors.  We note that an independent analysis of HST spectroscopy of SDSS1507+5230 led to a distance estimate of $250\pm50\pc$, \citep{Uthas2011} entirely consistent with both of our own values (Table 1).

For the determination of the X-ray luminosities carried out in \S4.3, we
adopted the distances based on our three-component model for the 16
systems with SDSS spectra, and the published distances for the
remaining four systems.

\vspace*{-0.5cm}

\subsection{Spectral analysis}
\label{s-xrayspectra}

In order to investigate the spectral nature of the twenty optically selected CVs we co-added their individual spectra using the {\rm ftool} \xselect. The combined spectrum was then grouped so as to provide a minimum of twenty counts per energy bin and thus enable the use of $\chi^{2}$ statistics.  Figure~\ref{f-spec} (\textit{left}) shows the average spectrum fitted with a simple one-temperature thermal plasma model (\mekal\ model in \xspec ; \citealt{mekel, mekel2}) having a temperature of $\approx6$\kev\ (Model~1; Table~2). This model provides a poor fit with $\chi^2/\nu = 31.3/19$ and exhibited residual at energies $\la 1\kev$.  A much improved fit was achieved by the addition of a second thermal plasma model \citep[in a similar manner to  e.g.][]{vanTeeseling1994, Wheatley1996} absorbed by a neutral hydrogen column (\wabs{\footnote{ {Using the standard BCMC cross-sections \citep{balucinska} and ANGR abundances \citep{abundances}.}}}; Model 2,  Fig.~\ref{f-spec}, \textit{right}), however, it was found that the value of \nh\ is degenerate with the plasma temperature. Nonetheless, for a typical value of \nh = $4\times10^{20}\pcmsq$ \citep{Baskill2005}, we find a temperature of $0.62^{+0.16}_{-0.24}\kev $,  consistent with the values found by \citet{Baskill2005} for a sample of 34 CVs using a similar model. We have thus fixed the value of \nh\ to $4\times10^{20}$cm$^{-2}$ when estimating the errors presented in Table~2.

The best fitting model for this sample of low-luminosity CVs is here  interpreted as a two collisionally-ionised plasma with temperatures of $8^{+10}_{-3}$ and $0.62^{+0.16}_{-0.24}$\kev. Of course, the fact that we have two temperatures does not necessarily imply  the presence of two discrete components in the X-ray spectrum  as this likely a simple approximation to common cooling flow models which successfully represent the spectra of more luminous CVs \citep[e.g.][]{Wheatley1996, Mukai2003, Baskill2005, Byckling2010}. The results presented here for the spectral properties of our sample of non-magnetic, low luminosity CVs are  again consistent with that of the Galactic ridge emission \citep[e.g.][]{Koyama1986, Kaneda1997}. Using the best-fit model, a count-rate of $4.0\pm0.2\times10^{-3} {\rm s}^{-1}$ resulted in a 0.5--10\kev\ unabsorbed energy flux\footnote{Unabsorbed flux obtained using the \xspec\ model \cflux\ convolved with Model 2 from Table~2.} of $1.6\pm0.2\times10^{-13}\ergpcmsqps$. These values were used as conversion factors between the 0.5--10\kev\ count rate and flux in the following section. We note here that for the only object common in both our sample and that of  \citet[][]{Byckling2010}  (namely ASAS0025+12), the luminosity found here of $L_{\rm 0.5-10} = (1.3\pm0.6) \times 10^{30}\ergps$ (Table~1), based on the methodology described above, is fully consistent with the value of $L_{\rm 2-10} = (1.6^{+3.8}_{-0.8})\times 10^{30}\ergps$ found in their work, even when considering the different energy range used in this work.

%ASAS0025+12 is the only object in common with the analysis of Byckling et al. 2010: they find L(2-10kev)=1.6+3.8-0.8x1e30 erg/s, which is actually very nicely consistent with your value, so I think it is worth mentioning this. 

\begin{table}
\begin{center}
  \caption{\label{t_fitx}Model parameters for the combined source spectra.}

\begin{tabular}{lcccccccccc}                
\hline
\hline
Parameter & Model 1 & Model 2 \\
\hline
\nh\ $(\times10^{20} \cm^{-2})$ & --   & 4      \\
$kT_1\kev$  &$5.8^{+2.0}_{-1.2}$  &  $8^{+10}_{-3}$      \\
$Norm_1$ ($\times10^{-5}$)  &  $8.2^{+0.8}_{-0.7}$  & $7.2\pm0.8$   \\ 
$kT_2\kev$  & --    &               $0.62^{+0.16}_{-0.24}$             \\
$Norm_2$ ($\times10^{-5}$) & -- &  $0.8\pm0.3$        \\
$\chi^{2}/\nu$        & $31.3/19$ & $11.3/17 $ \\
\hline
\hline
\end{tabular}
\end{center}

\small Notes-- Model 1 is the single-temperature thermal plasma model \mekal\ in \xspec. Model 2 assumes a two-temperature thermal plasma with photoelectric absorption (\wabs(\mekal+\mekal) in \xspec). The value of \nh\ was frozen at $4\times10^{20}$cm$^{-2}$. The quoted error corresponds to a 90 per cent confidence level for one parameter of interest ($\Delta\chi^{2}=2.71$). 

\end{table}

\begin{figure*}
\includegraphics[clip=true,scale=0.32] {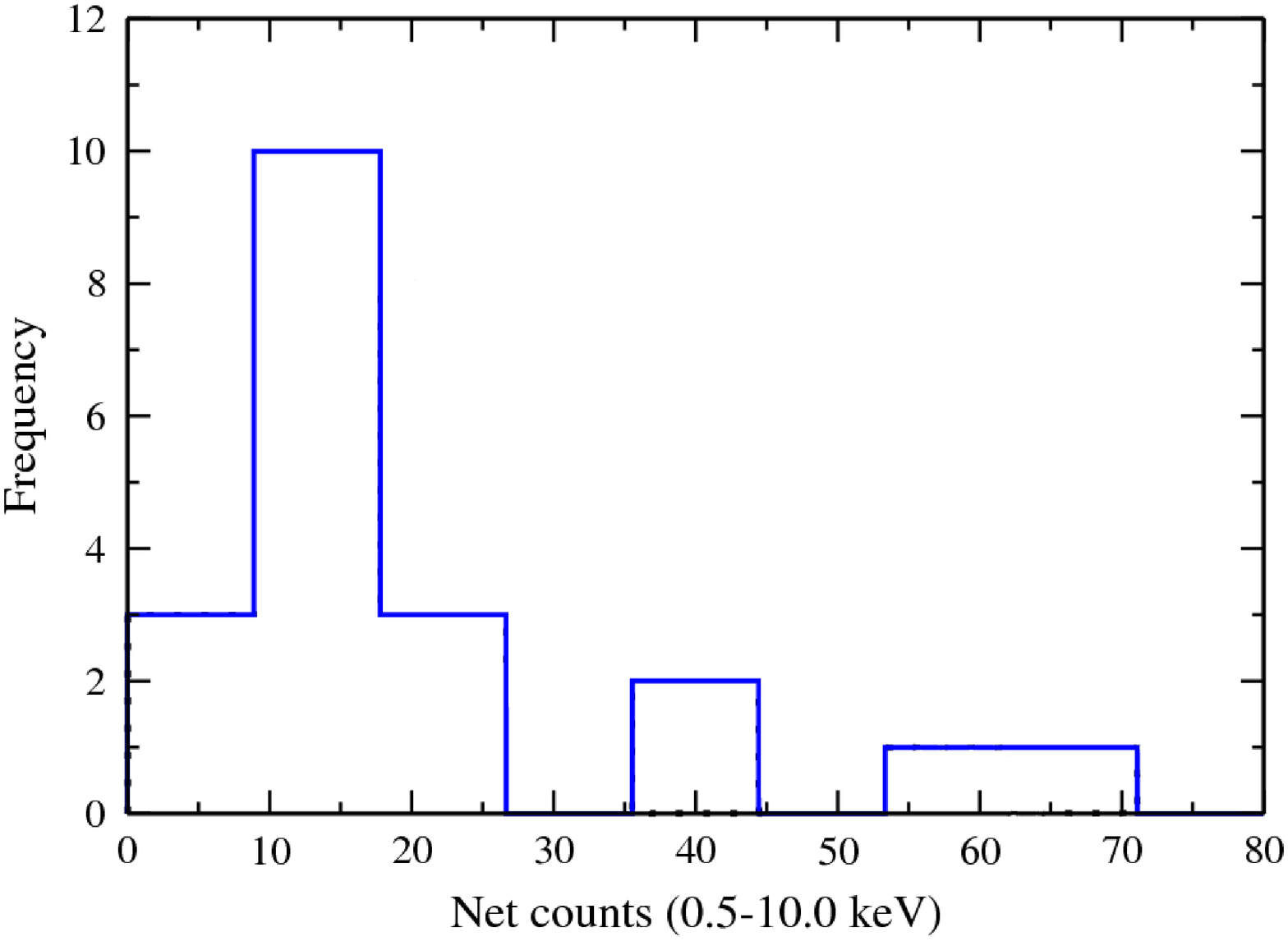}
\includegraphics[clip=true,scale=0.32] {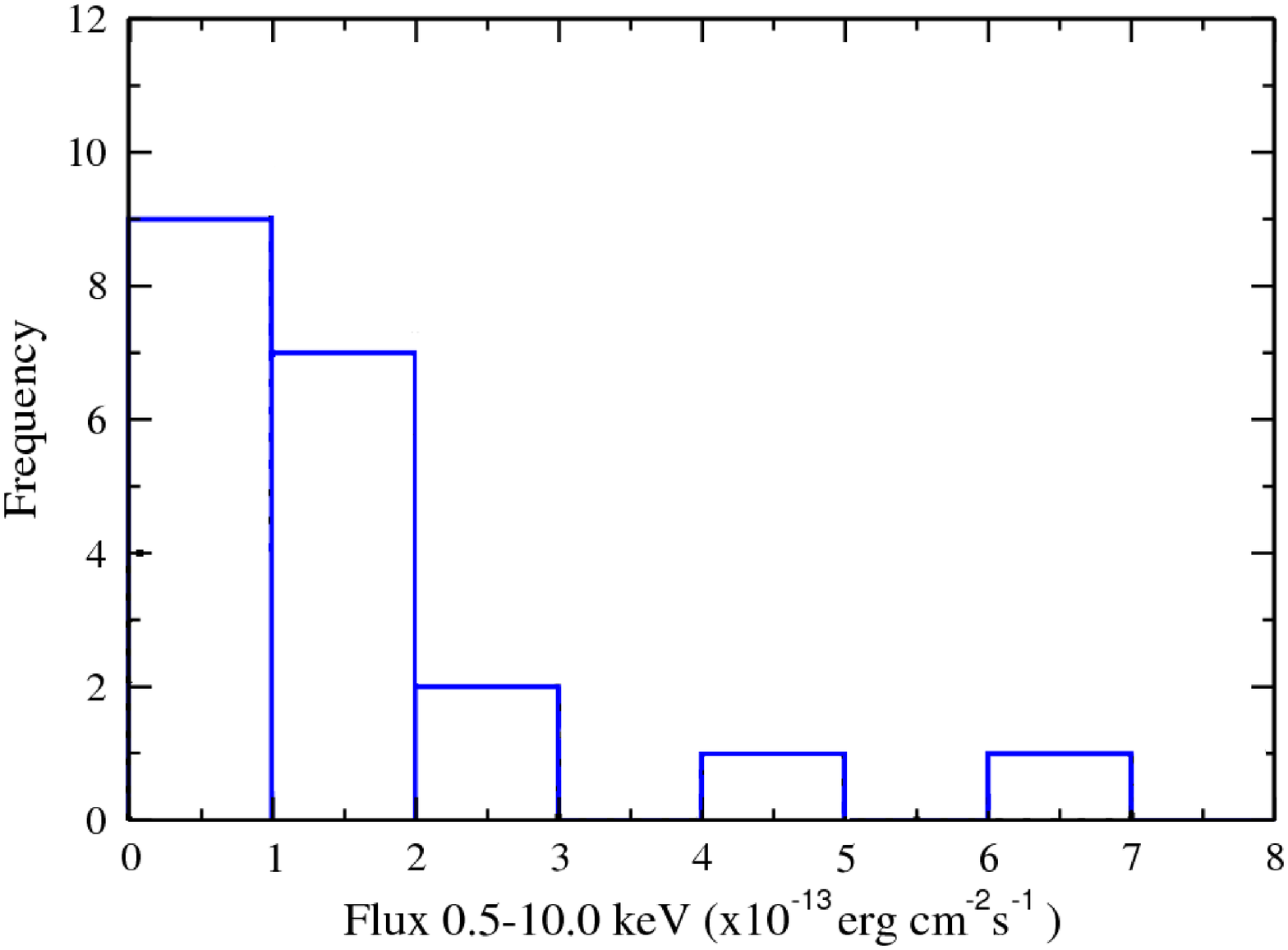}
\caption{{\it Left}:\label{f-his_cnts} Histogram of  counts detected in the energy band 0.5--10.0\kev. { \it Right}: Same as before but for the flux obtained using the rate--flux conversion factor as described in Sect.~\ref{s-xrayfluxes}. }
\end{figure*}

\subsection{X-ray photometry }
\label{s-xrayphotometry}

\subsubsection{Energy fluxes}
\label{s-xrayfluxes}
 A histogram of the number of sources as a function of the  counts in the 0.5--10.0\kev\ energy
range is shown in Fig.~\ref{f-his_cnts} (\textit{Left}). In order to obtain an estimate of the energy flux (in $\ergpcmsqps$) we used the spectral model described in the previous section to compute an average conversion factor between count-rate ({\it R}) and energy flux in the
range $0.5-10$\kev.  We find (Sect.\,\ref{s-xrayspectra}) that for the combined spectrum of all twenty sources, a two-temperature optically
thin plasma model  absorbed by a neutral
hydrogen column equivalent to \nh~$\approx4\times10^{20}$~cm$^{-2}$
yields a conversion factor for the unabsorbed energy flux of
$F_{(0.5-10)}=(4.1\pm0.4)\times10^{-11}\times R~\ergpcmsqps$, where {\it R} is the count-rate in the 0.5--10\kev\ energy band. As a consistency check we used  this model to fit the individual spectra of the 4 observations with over 40 counts (see Table~2) allowing only a multiplicative constant to vary. This resulted in values for the conversion factor varying from approximately $3.8$ to $4.6\times10^{-11}$, in agreement with the value above.  The fluxes for the various targets obtained as described are also listed in Table~1 and shown in Fig.~\ref{f-his_cnts} (\textit{Right}).

\subsubsection{Luminosity distribution}
\label{s-xrayluminosities}

 Table~1 lists the
calculated 0.5--10.0\kev\ luminosities of all twenty sources. The large error in these values are the result of propagated errors in both the distance and flux.  A
histogram of the number of sources as a function of log-luminosity is
displayed in Fig.~\ref{f-his_lum}. SDSS0843+2751 and SDSS0919+0857  were not used in the construction of the histogram due the sources having a count rate consistent with zero, possibly resulting from
their short exposure time (see Table~1). However the upper limits to
their luminosities are displayed with magenta and cyan arrows
respectively. The solid red line shows the Gaussian distribution for
the mean (0.5--10\kev) log-luminosity with $< {\rm  log} (L_{\rm 0.5-10}) > =
29.78$ and a variance of 0.16. For comparison, in Fig.~\ref{f-his_lum} we show in green the (renormalised) luminosity distribution obtained from the X-ray selected  \rosat\  all sky survey catalogue  from \citet{Verbunt1997} after scaling to the 0.5-10\kev\ range used here assuming  Model~2 (outlined in  Table~2 and described in \S~4.2).

\begin{figure}
\includegraphics[width=\columnwidth] {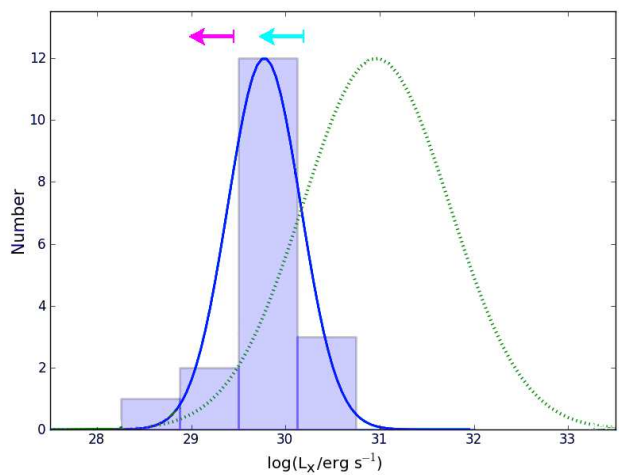}
\caption{\label{f-his_lum} X-ray luminosity distribution for a sample of optically-selected CVs (blue).  Only sources
  with observed fluxes consistently greater than zero are included in 
  the  histogram. The magenta and cyan arrows shows the upper limits
  for SDSS0843+2751 and SDSS0919+0857 respectively. The mean Gaussian distribution
  (solid blue) has $< {\rm  log} (L_{\rm 0.5-10}) > = 29.78$ with a variance of 0.16. For comparison with previous, X-ray selected surveys, we show in  green the (renormalised) Gaussian distribution obtained for the 46 sources catalogued by the \rosat\ all sky survey (derived from \citealt{Verbunt1997} scaled to the 0.5-10\kev\ range used in this work.  This illustrative, X-ray selected, distribution has a  mean value of $< {\rm  log} (L_{\rm 2-10}) > = 30.96$ and a variance of 0.78, clearly higher than our sample.   }
\end{figure}

\vspace*{-0.5cm}

\section{Discussion}

\begin{figure*}
\vspace{-0.cm}
{\includegraphics[height=7.8cm] {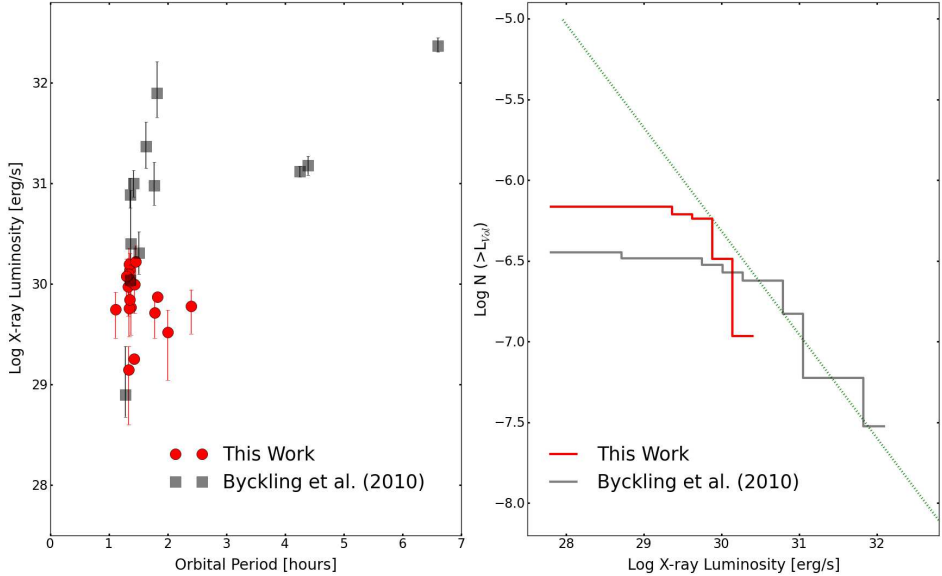}}
\vspace{-0.2cm}

\caption{\textit{Left:} X-ray luminosity (0.5-10\kev) versus orbital periods for the sample presented in this work (red) and that of \citet[][their Fig.~7]{Byckling2010} (black). The 2-10\kev\ flux quoted in the latter was converted to the 0.5-10\kev\ range using the model detailed in Table~2. \textit{Right:} Cumulative source distribution as a function of X-ray luminosity for our sample (red) and for that  presented in  \citet[][their Fig.~6]{Byckling2010} (black). The dashed line shows the powerlaw $N(>L) = k(L/L_t)^{- \alpha}$, where $k = 2.39\times 10^{-7} {\rm pc}^{-3}$ and $L_t=3\times 10^{30}\ergps$ as per \citet[][]{Byckling2010}. } 
\end{figure*} 

Suggestions that the Galactic ridge X-ray emission is diffuse came about due to failure in fully resolving the emission into point sources \citep[e.g.][]{Ebisawa2001Sci}.  Non-magnetic CVs have also been dismissed as a main contributor to the GRXE \citep[e.g.][]{Sazonov2006, Revnivtsev2006grxe}  due to suggestions that the overall population is too luminous, with a mean  X-ray luminosity greater than $\sim10^{31}\ergps$ \citep[e.g.][]{Eracleous1991, Richman1996,  Verbunt1997,  Baskill2005}. We argue here that the latter is likely a result of a selection bias in their X-ray luminosity function caused by the predominant use of X-ray selected samples. By studying an optically-selected sample of  non-magnetic, intrinsically faint   CVs, we find a population of low X-ray luminosity CVs, having an average luminosity of  $< L_{\rm 0.5-10}>=8\times10^{29}\ergps$ which still resembles the X-ray spectrum of the Galactic Ridge. This population, if found to exist in large numbers, could be a strong  contributor to the overall GRXE below $\sim10$\kev.

The X-ray spectra of  CVs are often characterised by a two-temperature plasma model \citep[see e.g.][]{Mukai1993, Baskill2005}. In Sect.\,~\ref{s-xrayspectra} we showed that the combined spectrum of this sample of twenty low-luminosity CVs has the same spectral characteristics as that of the better known, high-luminosity CVs (Fig.~\ref{f-spec}). This two-temperature spectrum  closely resembles that of the Galactic ridge X-ray emission \citep{Mukai1993}.  Eighteen out of the twenty sources in our sample were detected with Swift XRT and were found to have X-ray luminosities in the range $10^{28-30}\ergps$. Figure~\ref{f-his_lum} shows this distribution together with the upper limit of the two sources having a count rate consistent with zero. The peak of the distribution for the optically-selected sample used here is considerably lower than that presented by \citet{Verbunt1997} of $<{\rm log}(L_{\rm 2.0-10})>= 30.8$ (shown in green in Fig.~\ref{f-his_lum}), especially after  considering the larger energy range used here (Fig.~3). This difference is clearly a result of the different biases in the selection of the two  samples. 

\citet{Byckling2010}  presented a thorough  analysis of the X-ray luminosity function in an optically-selected sample of CVs  having parallax-based distance measurements. In that study,  the authors  find a peak in the luminosity function at $\sim10^{30-31}\ergps$, systematically lower than previous -- X-ray selected -- studies but still more luminous than the result found here.   Figure~5 (\textit{left}) shows the \citet{Byckling2010} sample (black) together with our results (red) for the Luminosity--Orbital period plane.   It is clear that we are  probing a different parameter space to \citet{Byckling2010}, with overall low orbital periods as expected for more evolved, and thus intrinsically fainter, systems. We also show in Fig.~5 (\textit{right}) the cumulative distribution -- Log~N ($>L_{\rm Vol}$) where $L_{\rm Vol}$ is the luminosity per cubic parsec volume as a function of Log~L.  For the sample of \citet{Byckling2010}, we assume a volume out to 200~pc, in accordance to the previous authors. For the sample presented here, the furthest object is located  $\sim400$~pc away (Table~2). However, as our sample is not complete out to this radius, we estimate a correction factor by noting that SDSS covered 8,000$\degsq$ ($\sim20\%$; see \S 2) of the sky and our sample uses 16 out of the 30 low luminosity systems detected in DR5. As the size of the population presented here is still highly uncertain, the normalisation of the logN--logS plot, and with it the space density of these systems, should be interpreted with extreme caution. Nonetheless, these results show that the sample presented in this work  fills the gap between the only low-luminosity outlier (GW Lib at $L_{\rm 2-10} \sim5\times10^{28}\ergps$) in the sample of \citet{Byckling2010}, and the peak of their distribution at $\sim10^{30-31}\ergps$.

It was shown by \citet{Pretorius2007} and  \citet{Pretorius2012}  that the vast majority of CVs could be fainter than $\sim5\times10^{29}\ergps$ -- and thus yet undetected -- based on the  limits to their CV space density as found from \rosat\ surveys. \citet{Muno2004} showed that to account for the diffuse emission within 20\pc\ of the Galactic centre, approximately $0.2\%$ of stellar mass would have to be hard X-ray sources with $L_{\rm x}>3\times10^{29}\ergps$. The total stellar mass within 20\pc\ of the Galaxy centre is estimated as $\approx10^{8} \msun$ \citep{Launhardt2002} and hence $\approx2\times10^5$ sources would be needed in that region. This is equivalent to a source density of $\sim6\pc^{-3}$ assuming a spherical region of radius 20\pc. By scaling according to the mass model of \citet{Launhardt2002} with a stellar density in the inner 20\pc\ of $1000\msun\pc^{-3}$ and a local stellar density of  $0.1\msun\pc^{-3}$ the source density in the 20\pc\ region corresponds to a local density of $\approx6\times10^{-4}\pc^{-3}$, which is significantly higher than that found in most current observational campaigns. However, population synthesis based on the standard CV evolution scenario indeed predict space densities which are larger than the current observed values, ranging from $\sim10^{-5}\pc^{-3}$ \citep{Politano1996} to a few  $\sim10^{-4}\pc^{-3}$ \citep{Kool1992}. These models further predict that the population should be dominated by short period systems (see e.g. \citealt{Kolb1993, Kolb1996}), similar to the population presented in this work (see Fig.~5). Interestingly, the upper limit on  low-luminosity CVs -- the undetected population in \citet{Pretorius2012} -- does indeed add further empirical credence to population syntheses models (see Fig.~4 of  \citealt{Pretorius2012}). If population syntheses models are correct then the population density for CVs should be greater than observed, and CVs having  luminosities of a few $10^{29}\ergps$ could thus be a major contributor to the GRXE. 

It is now established that intermediate polars (IP) are important contributor to the GRXE above $L_{\rm x} \sim 10^{30}\ergps$ \citep[e.g.][and references therein]{Hong2012}. However, IPs are both theoretically and empirically known to belong to the brightest and rarest end of the CV population. In a volume limited sample, IPs represent at best a few per cent the entire CV population \citep{DownesCV2001, DownesCV2006}.

 %with  approximately 1 IP for \textit{at least} every 100 known non-magnetic, fainter CVs  similar to the population described here.{\footnote{ As the census for magnetic CVs are more complete compared to their non-magnetic counterparts due to them being more luminous, the true ratio between IPs and non-magnetic CVs may be much more extreme.}.

\vspace*{-0.5cm}
\section{Conclusions}

Using \swift -XRT, a sample of twenty optically selected, non-magnetic CVs were observed and their X-ray luminosities (0.5-10.0 keV) determined. The sources luminosities were found to range between  $10^{28}$ to $10^{31}\ergps$ with an average value of $8\times10^{29}\ergps$. Cataclysmic variables towards the low luminosity range would not have been detected by the various \chandra\ surveys of the Galactic centre, where their limiting sensitivity is  $L_{\rm 0.5-7keV}\sim10^{30}\ergps$ at a distance of 8\kpc\ \citep{Revnivtsev2009}. Spectral analyses were made of the average co-added spectrum. The result is consistent with a two-temperature plasma at $\sim0.6\kev$ and
$\sim8\kev$, similarly to that of the well studied high-luminosity population. The similarity between the spectra of the point sources and that of the GRXE suggests that this
emission could indeed be produced by CVs. The present work demonstrates the use of optical selection in order to remove X-ray luminosity bias in CVs. The study of a larger optically selected sample would be advantageous in better understanding the luminosity function of CVs, however in order to determine the absolute contribution of CVs towards the GRXE one also needs to determine the corresponding population density function.  

\vspace*{-0cm}
\section*{Acknowledgements}

RCR  thanks the Michigan Society of Fellows and NASA.
RCR is supported by NASA through the Einstein Fellowship
Program, grant number PF1-120087 and is a member of the
Michigan Society of Fellows. PJW and BTG acknowledge support
from the UK STFC in the form of a Rolling Grant. JPO acknowledges the support of the UK Space Agency. This work also made use of data supplied by
the UK Swift Science Data Centre at the University of Leicester. We also thank the anonymous referee for his/her constructive comments.

\bibliographystyle{mnras}                                              
\bibliography{ref.bib}

\end{document}